# Generating Exact- and Ranked Partially-Matched Answers to Questions in Advertisements

Rani Qumsiyeh
Computer Science Dept.
Brigham Young University
Provo, Utah 84602, U.S.A.
rani_qumsiyeh@yahoo.com

Maria S. Pera
Computer Science Dept.
Brigham Young University
Provo, Utah 84602, U.S.A.
mpera@cs.byu.edu

Yiu-Kai Ng
Computer Science Dept.
Brigham Young University
Provo, Utah 84602, U.S.A.
ng@compsci.byu.edu

## ABSTRACT

Taking advantage of the Web, many advertisements (ads for short) websites, which aspire to increase client's transactions and thus profits, offer searching tools which allow users to (i) post keyword queries to capture their information needs or (ii) invoke form-based interfaces to create queries by selecting search options, such as a price range, filled-in entries, check boxes, or drop-down menus. These search mechanisms, however, are inadequate, since they cannot be used to specify a natural-language query with rich syntactic and semantic content, which can only be handled by a question answering (QA) system. Furthermore, existing ads websites are incapable of evaluating arbitrary Boolean queries or retrieving partially-matched answers that might be of interest to the user whenever a user's search yields only a few or no results at all. In solving these problems, we present a QA system for ads, called CQAds, which (i) allows users to post a natural-language question $Q$ for retrieving relevant ads, if they exist, (ii) identifies ads as answers that partially-match the requested information expressed in $Q$, if insufficient or no answers to $Q$ can be retrieved, which are ordered using a *similarity-ranking* approach, and (iii) analyzes incomplete or ambiguous questions to perform the "best guess" in retrieving answers that "best match" the selection criteria specified in $Q$. CQAds is also equipped with a Boolean model to evaluate Boolean operators that are either *explicitly* or *implicitly* specified in $Q$, i.e., with or without Boolean operators specified by the users, respectively. CQAds is easy to use, scalable to all ads domains, and more powerful than search tools provided by existing ads websites, since its query-processing strategy retrieves relevant ads of higher quality and quantity. We have verified the accuracy of CQAds in retrieving ads on eight ads domains and compared its ranking strategy with other well-known ranking approaches.

## 1. INTRODUCTION

There are huge collections of (un-/semi-)structured data on the web these days which hold diverse information, including advertisements (ads for short), such as car ads, job ads, house-for-sale ads, rental ads, etc. The web is a perfect publication forum for ads, since ads websites (allow sellers to) post ads for potential buyers who can freely access archived and newly-created ads anytime and anywhere, which cannot be achieved by any traditional publication media. Many ads websites, such as ebay.com, provide an easy-to-use interface that allows users to create keyword queries for retrieving ads of interest or specify their search criteria using pre-defined check boxes and push-down menus to simplify the query construction process. The search mechanisms employed by these ads websites to identify ads that satisfy users' needs, however, are inadequate and insufficient, since a number of ads queries can only be formulated using rich syntactic- and semantic-structure, such as "Find single-family units that are not condos, with price no more than $300k, and located between West Jordan and Bountiful, Utah". To solve this problem, we present a closed-domain question-answering (QA) system on ads, denoted CQAds, which can answer any ads questions. Unlike existing QA systems, CQAds retrieves answers to a question $Q$ without the burden of performing syntactic analysis on $Q$. It interprets the semantics of $Q$ at most by employing the simple context-switching analysis, which facilitates the process of transforming the information needs expressed in $Q$ into a SQL query to be evaluated against the underlying ads data (records), which yield the source of answers to $Q$. Moreover, if $Q$ includes (implicit or explicit) *Boolean* operators, CQAds evaluates $Q$ using a novel Boolean model, which *infers* Boolean operators, if necessary, to handle the problem of contradictory searching criteria specified in $Q$ and determines the evaluation order of the selection criteria. Furthermore, whenever exact-matched answers to $Q$ are lacking, CQAds retrieves answers that *partially match* the information need expressed in $Q$ ordered according to a *similarity-ranking* approach based on various domain-specific correlation measures.

To enhance the effectiveness and efficiency of processing ads questions, CQAds (i) first uses the Naive Bayes classifier based on the Joint Beta Binomial Sampling Model to automatically identify the domain of an ads question $Q$, (ii) corrects spelling mistakes using a trie when evaluating $Q$, (iii) performs an intelligent "best guess" and retrieves answers that most adequately match an incomplete and/or ambiguous selection criterion specified in $Q$, (iv) applies substring matching to speed up the process of retrieving answers to $Q$, and (v) processes each SQL query converted from $Q$.

CQAds offers a fully-automated, powerful question-answering search engine to help its users locate desired ads of interest. CQAds does *not* require experts to build complex ontologies or knowledge-based systems to interpret and retrieve answers to users' questions; instead, it solely relies on a set of simple, pre-defined rules that transform a question into its corresponding SQL statement. It employs a Boolean model and question-processing techniques to handle explicit and implicit Boolean queries that enrich the *expressive power* of CQAds. Moreover, CQAds retrieves partially matched answers to a user's question using a *similarity* formula, which relies on numerical attribute value proximity, as well as categorical attribute value similarity based on a novel query-log analysis approach. CQAds introduces new and novel methods for query-sense disambiguation, query interpretation, incomplete questions analysis, and shorthand notations identification on users' questions.

We organize our paper as follows. In Section 2, we discuss previous work on closed-domain QA systems and existing strategies on ranking partially-matched answers. In Section 3, we introduce the Naive Bayes classifier used for classifying users' questions into their corresponding ads domains. In Section 4, we detail the design of CQAds for processing (Boolean) ads questions. In Section 5, we present the experimental results which verify the effectiveness and efficiency of CQAds in answering questions on eight distinct ads domains. In Section 6, we give a conclusion. Throughout the paper, we demonstrate CQAds using car ads as a running example.

## 2. RELATED WORK

We present existing closed-domain QA systems and different ranking strategies in this section.

### 2.1 Closed-domain QA systems

Chung et al. [4] propose a weather forecasting QA system based on a named-entity tagger, a dependency parser, a set of inference rules, and an ontology that distinguishes weather terms and information about cities. Wang [22] also relies on an ontology and predefined question templates to answer questions in mobile-service areas. Vargas-Vera and Lytras [21] introduce AQUA, a framework that combines natural language processing, ontologies, logic, and information retrieval technologies to answer questions on academic people and organizations. Ferrandez et al. [8] develop QACID, an ontology-based question answering system applied to the cinema domain, which depends on (i) a domain ontology, (ii) cinema-related information, (iii) a lexicon that matches words in natural language queries and ontology instances, and (iv) a pattern database (DB) to deduct semantic inferences between a query Q and its patterns. Contrary to CQAds, the systems in [4, 8, 21, 22] depend on ontologies that must be manually constructed to answer questions. Furthermore, these QA systems do not provide a partial matching mechanism, which retrieves closely-related results if no exact-matched answers exist.

A medical QA system is introduced in [20]. Given a user's question $Q$, the QA system relies on WordNet, Unified Medical Language System (a previously developed medical question taxonomy), and natural language processing techniques, to retrieve passages from a set of documents that likely contain the answers to $Q$. The system in [20] retrieves only exact-matched answers.

Wang and Luo [23] present a Chinese QA system that answers questions in the telecom product domain. The authors assume that a user's query $Q$, as well as web documents with the answers to $Q$, are semantically well-formed. However, since there is no central editorial board to edit documents on the web, the semantics and grammar of sentences in a web page are not always well-formed, which contradict the assumption made by the authors. Moreover, the system in [23] cannot handle short, poorly-formed questions.

### 2.2 Ranking partially-matched answers

Most ranking approaches, which can be applied to QA systems, focus on defining a *scoring function*, ranging from using the Euclidean distance to the probabilistic model [12] to quantify the degree of relevance between a query and a potential answer. These scoring functions, however, are not effective for every domain, since a partially-matched answer $A$ to a query $Q$ does not necessarily mean that keywords in $Q$ are highly similar to keywords in $A$. For example, "Honda Accord" is relevant to a search for "Toyota Camry", since both cars are compact and offer similar features. CQAds, on the other hand, relies on word similarity, in addition to other similarity measures not based on (exact) keyword matching, to retrieve partially-matched answers to a user's question.

Das et al. [6] introduce an attribute-selection method that determines the top-$m$ attributes which the users are most likely interested in for ranking the top-$n$ retrieved results of a DB query. The main difference between CQAds and the approach in [6] is their complexity. The "$N$-1" approach of CQAds selects the attributes to be considered and ranks retrieved records in a simple linear order.

In [10], the SQL query language is extended to allow users to indicate the preferred ranking function on the attributes specified in a DB query. The ranking strategy, however, relies on user's feedback that in turn requires background knowledge in using SQL, which ordinary users are lacking. CQAds does not require any human feedback in processing a user's question.

Bilotti et al. [2] require the user to specify semantic constraints on his/her query in order to determine the relative importance of different selection criteria included in the query, which are used to build structured queries to retrieve answers. Although effective, the proposed method is not robust for ranking partial matches and requires additional user's input for specifying the semantic constraints. Pizzato et al. [18] apply the vector-space model (VSM) to represent words in a question and their semantic role labels to rank answers in response to a user query. This model does not capture linguistic and semantic relationships between words, which results in a low quality and quantity of partially matched answers.

FAQFinder [3] uses a VSM engine to retrieve a list of FAQs relevant to a user's question $Q$. When $Q$ is submitted, FAQFinder uses the keywords in $Q$ to find a question-answer pair among the FAQs that best matches $Q$, which is similar to CQAds in retrieving partially-matched, related answers to a question, if they are needed. FAQFinder, however, is neither scalable nor reliable, since to answer $Q$, a question-answer pair that closely resembles $Q$ must exist in order to provide the user with an answer to $Q$.

## 3. CLASSIFYING USERS' QUESTIONS

One of the major tasks of CQAds is to identify the ads domain, such as Cars-for-Sale domain, to which a user's question belongs. Manually determining the ads domain of the user's question is not feasible, which may require the user's intervention and is a tedious and inefficient process. In solving this problem, CQAds relies on the Naive Bayes classifier based on the Joint Beta Binomial Sampling Model (JBBSM) to identify the domains of users' questions.

The Naive Bayes classifier is simple, easy to implement, robust, highly scalable, and domain independent. The classifier relies on the conditional distribution of the class variable to compute the probability of assigning the natural class $c$, i.e., an ads domain in our case, to a document $d$, i.e., a user's question in our case, which is the well-known Bayes' Theorem.

$$P(c \mid d) = \frac{P(c)P(d \mid c)}{P(d)} \quad (1)$$

where $P(d)$ is the probability of a given document $d$, $P(c)$ is the probability of a particular natural class $c$, and $P(d \mid c)$ is the probability of $d$ given that class $c$ occurs.

In choosing the ads domain $c$ to which user's question $d$ should be assigned, we compute the conditional probability $P(c \mid d)$ as defined in Equation 1 for each one of the possible ads domains. We assign to $d$ the ads domain $c$ that yields the highest $P(c \mid d)$ among all the ads domains (as shown in Equation 2).

$$Class(d) = argmax_{c \in C} \ P(c \mid d) \qquad (2)$$

where $C$ is the set of ads domains.

In estimating $P(d \mid c)$ in Equation 1, we have chosen JBBSM [1], which considers the "burstiness" of a keyword, i.e., a keyword is more likely to occur again in $d$ if it has already appeared once in $d$. JBBSM accounts for unseen words in a document.

## 4. OUR QUESTION-ANSWERING SYSTEM

In this section, we introduce the question processing and evaluation strategies of CQAds on ads.

### 4.1 Ads data and question pre-processing steps

CQAds, which relies on a DB that archives ads in different domains (with a table in the DB for each domain) for answering ads questions, has been designed independently of its underlying ads DB. In its current implementation, CQAds depends on the tool introduced in [17] to create the underlying DB. In addition, the relational schema for each ads domain in the DB has been predefined based on the attribute values and column names specified in ebay.com, which is the largest, most comprehensive e-commerce website on the web.

In a DB, data records are defined using fundamentally different types of attributes, along with their domain values as specified in a DB schema. There is a natural difference between a *categorical attribute value* (such as a car model) and a *numerical value* (such as a price) in the same record. There are also differences in terms of how users specify search criteria on these two types of data. For example, in searching for a particular car the users invariably include the $Make$ and $Model$ in a question; however, for the price, they may specify a range with(out) a dollar sign ('$'), or may simply be seeking the cheapest ones. To handle the inherent variance in expressing information needs specified by the users in natural language ads questions, CQAds identifies the selection criteria specified in each question by labeling them with their corresponding data types (presented in Section 4.1.1), which are eventually used to match attribute values in DB records.

#### 4.1.1 Attribute types in the ads DB

Each ads record in the underlying DB showcases a particular product or service $PS$, which can be recognized by its *unique identifier* formed by one or more attribute values in the ad. In addition, each ad often includes (i) a number of *properties* that describe $PS$ and/or (ii) *quantitative values* that identify the measurable substances of $PS$. With that in mind, we have defined the following ads *data types*, of which the corresponding data items are alpha-numerical strings.

*Type I* attribute values in an ad, which showcases $PS$, yield the *unique identifier* of $PS$ that are required values to be included in an ad (its DB record, respectively). Type I attributes are *primary-indexed* fields of the relational schema which defines the corresponding ads domain. Sample Type I attributes in the Cars-for-Sale ads domain are "Maker" and "Model", and "Toyota" and "Camry" are one of their respective values.

*Type II* attribute values describe the *properties* of $PS$ in an ad, which are not required values in the ad that showcases $PS$. Type II attributes are *secondary-indexed* fields in the corresponding relational schema. "Color" is a Type II attribute in the Car-for-Sale ads domain, and "Blue" is one of its domain values.

*Type III* attribute values specify the *quantitative values* of $PS$ in an ad. A sample *Type III* attribute is "Salary" in the Jobs ads domain, and $50,000 is one of its values. In addition, "usd" is also a *Type III* attribute value, which identifies the unit of "Price" (a *Type III* attribute) in the Cars-for-Sale ads domain.

#### 4.1.2 Conditions specified in users' ads questions

Any constraint on an attribute value a user specified in an ads question constitutes a *condition*. Examples of conditions specified in a car-ad question that include Type I attribute values are 'Honda' ($Make$) and 'Accord' ($Model$), Type II are 'automatic' ($Transmission$) and '4-wheel drive' ($DriveTrain$), and Type III are 2004 ($Year$) and less than $5000 ($Price$). In designing CQAds, we assume that a user's question can include conditions that involve any combination of Types I, II, and III attribute values.

EXAMPLE 1. Listed below are a few sample car-ads questions.

$Q_1$: Do you have a *2 door red BMW*?

$Q_2$: *Cheapest 2dr mazda* with *automatic* transmission

$Q_3$: I want a *4 wheel drive* with *less than 20K miles* □

To process an ads question with conditions specified on Types I and II attribute values, CQAds simply *matches* (i.e., performs equality comparisons of) the search values (as conditions) to the corresponding ones in a DB record. Evaluating conditions on Type III attribute values, however, is more involved.

User-specified selection criteria on Type III attribute values can be categorized as either *range* or *exact matches*, such that (i) users may seek ads *above, below*, or *between* certain values, or (ii) they may seek ads with *extreme* values within a selected group, such as "the cheapest car". CQAds handles both cases by defining what constitutes a *complete condition* versus *partial condition*. *Context-switching analysis* is applied to identify partial conditions and merge them with proximity keywords, if they exist, in a question to create *complete conditions*. These analyses are based on superlatives and boundaries, which are defined below.

- *Superlatives* ($S$): used for finding ads with $max/min$ values

  - *Complete* ($C$): Stand-alone terms that ask for extreme values, such as $cheapest$, $newest$, and $oldest$

  - *Partial* ($P$): Query terms that compare *extreme* values, e.g., $fewest$, $greatest$, $highest$, $least$, $lowest$, $max$, and $min$

- *Boundaries* ($B$): used for finding ads fitting a certain *range*

  - *Complete* (C): Query terms that compare numerical attribute *values*, e.g., *cheaper/less (or more) expensive*, and *newer/older than*

  - *Partial* (P): Query terms that require the specification of an *attribute* and one of its *values* for comparisons, e.g., *above, between, below, greater/higher/less/lower/more than, within*, and *under*

#### 4.1.3 Keyword tagging using a trie

Given a user's ads question $Q$, CQAds first $tags$ keywords (i.e., attribute values) in $Q$ with the appropriate labels to interpret the information need specified in $Q$, which in turn are translated into a SQL query statement to be evaluated against the underlying DB records. In order to automate the keyword tagging process, we use a *trie*, each of which is created for a distinct ads domain to parse and tag the keywords in a(n) (incomplete) user's question.

A trie, which is an ordered tree data structure, is an ideal choice for string processing, since it is faster and saves more disk space than other data structures. Using a trie, searching for a string (i.e., word) of length $m$ takes $\mathcal{O}(m)$ time, as compared with a *binary search tree* which requires $\mathcal{O}(m \log n)$ time, where $n$ is the number of elements in the tree. Tries are also better than *hash tables*, especially when the number of items to be searched is relatively small and static, which is the case in CQAds. The size of each trie used by CQAds is less than 50 megabytes.

Since CQAds translates each user's question $Q$ into an SQL query for processing, it must (i) *identify* all the *selection criteria* specified in $Q$ and their corresponding attribute value types, i.e., Types I, II, or III, (ii) *translate* each one independently into an SQL subexpression, and (iii) *combine* them into an SQL query. There are two types of keywords a user can include in a question processed by CQAds, i.e., the stand-alone and combined keywords. *Stand-alone* keywords in a question exist independently, and each conveys a selection condition by itself, such as the $Make$ or $Model$ of a car. Each stand-alone keyword $K$ is recognized by a trie when there exists a path from the root to a leaf node in the trie when $K$ is parsed. $Combined$ keywords in a question, however, do not convey a selection condition independently, i.e., they require other identifying keyword(s), such as $Mileage$ and $dollars$ ($), and their corresponding values. Combined keywords $CK$ can be detected by locating a space after a node in the trie labeled by the first keyword in $CK$ and the remaining path to a leaf node in the trie yields the remaining keywords in $CK$. Each node $N$ in the trie has a value and a label. The *value* of $N$ is the letter $N$ represents, whereas the label of $N$ is the *concatenation* of all letters on the path from the root node to $N$. If the label of a node is a valid English word (identified by online dictionaries), the node is labeled by a keyword.

Every node $N$ labeled by a keyword $K$ in a trie is assigned an *identifier* (ID), which is an interpretation of the functionality of $K$. Identifiers are used by CQAds for translating a user's question into its equivalent SQL query and are pre-programmed into the trie. A node identifier is determined based on (previous) node identifiers in the path from the root node of the trie. When a leaf node $N$ with an identifier $I$ is encountered, CQAds adds $I$ to a list, which maintains all the identifiers in the order they are detected. If $N$ is not a leaf node, then the next keyword in the question is scanned; the process is repeated until either a leaf node is reached or the potential missing identifier is detected (in a complete ads question.)

### 4.1.4 Trie construction

A trie is built for each ads domain in CQAds. To construct a trie for an ads domain $D$, CQAds requires the relational schema for $D$ and the trie identifiers, which are pre-defined, and the identifiers are the same for each ads domain. The identifiers used in the tries are shown in Table 1, called *identifiers table*.

Prior to constructing the trie for the first ads domain, the *identifiers table* is created manually, which is used by all the tries of different ads domains. The table includes a list of *keywords*, which are either comparison operators or phrases "Type I/II/III attribute values", and an identifier is assigned to each keyword. Using the identifiers table, if a given keyword $K$ is a Type I attribute value, then CQAds assigns $K$ the identifier ' "Type I attribute" = $K$'.

To add a new ads domain $D$ to CQAds, we extract from multiple ads websites a set of randomly-selected 500 ads belonged to $D$. We (i) manually create an ads *domain-specific table* for $D$, which includes the Type I attribute values extracted from the push-down menus provided by the ads websites and Types II and III attribute values in the ads, and (ii) construct the *trie* for $D$. Each character $C$ in each non-stopword in the ads is sequentially added as a node $N$ in the trie such that the *value* of $N$ is '$C$' and the *label* of $N$ is the concatenation of the values of the nodes preceding $N$. If the label of $N$ is a valid entry, a comparison operation (an attribute value, respectively), which is included in the identifiers table (the corresponding domain-specific table, respectively), $N$ is assigned the $ID$ extracted from the corresponding entry in the identifiers table. Otherwise, a new node $N$' is created for the character following $C$ and $N$' is linked to $N$.

During the process of identifying the types of keywords in an ads question $Q$ of domain $D$ using the corresponding trie and prior to evaluating $Q$ against the DB records of $D$, CQAds eliminates all the *non-essential* keywords, which are (i) stopwords, which carry little meaning, and (ii) keywords in $Q$ that are neither superlatives/boundaries (as defined in Section 4.1.2) nor Type I/II/III attribute values in $D$.

EXAMPLE 2. Consider the sample questions in Example 1 again. After non-essential keywords are removed from each of the questions, the simplified questions are:

$Q_1$: *2 door red BMW*

$Q_2$: *Cheapest 2dr mazda automatic*

$Q_3$: *4 wheel drive less than 20k miles*

Subsequently, the identified essential keywords, along with their tagged corresponding types and/or selection conditions, in the simplified questions are shown below.

$Q_1$: *"2 door"/TII "red"/TII "BMW"/TI*

$Q_2$: *"Cheapest"/TIII-CS "2dr"/TII "mazda"/TI "automatic"/TII*

$Q_3$: *"4 wheel drive"/TII "less than"/TIII-PB "20k mi."/TIII-CB*

where $TI, TII, TIII, C, P, S$, and $B$ are defined as in Sections 4.1.1 and 4.1.2, and $T$ stands for *Type*. □

The importance of properly identifying Type III elements is evident from the evaluation order discussed in Section 4.3.

## 4.2 Misspellings, Incomplete Queries, and Shorthand Notations

Users may create questions with errors. We have implemented simple algorithms to handle *spelling mistakes* (discussed in Section 4.2.1), *missing attributes* (detailed in Section 4.2.2), and *shorthand* notations (introduced in Section 4.2.3) in users' questions.

### 4.2.1 Using a prefix trie to correct spelling

Occasionally, a user *misspells* a word or forgets to add *spaces* in between keywords when posting a question $Q$. For example, a user might create the question, "Hondaaccord less than $2000" or "honda accorr less than $2000". To enhance the *effectiveness* and *user-friendliness* of CQAds, such errors are detected and corrected automatically using the trie of the domain to which $Q$ belongs, instead of returning irrelevant or no results to the user.

During the process of parsing a user's question $Q$, CQAds scans through each keyword $K$ in $Q$ by reading its letters one by one. If an end of a branch in the trie is encountered and no more characters are left in $K$, CQAds treats $K$ as a *valid* keyword; otherwise, if there are more characters left in $K$, a space is inserted at the current position of $K$, assuming that the user has forgotten to add a space. However, if $K$ is not recognized by the trie (i.e., none of the next letters in the trie matches the next letter in $K$), then $K$ is treated as a *misspelled* word $W$. CQAds compares $W$ with the alternative keywords recognized by the trie, starting from the current node in the trie where $W$ is encountered, using the "$similar\_text$" function which calculates their similarity based on the number of common characters and their corresponding positions in the strings. $Similar\_text$ returns the degree of similarity of two strings as a *percentage*. CQAds replaces the misspelled keyword in $Q$ by the alternative one with the *highest* similarity percentage.

**Table 1: Identifiers used by tries for tagging keywords**

| Keyword | Identifier (ID) |
|---|---|
| A Type I attr. value | ID := ' "Type I attr." = keyword' |
| A Type II attr. value | ID := ' "Type II attr." = keyword' |
| A Type III attr. value (keyword $K$) | If ID == '', then ID :=' "Type III attr. $K$" ' <br> If ID =='group by', then <br>   ID := 'group by "Type III attr. $K$" ' <br> If ID == '<', ID := ' "Type III attr. $K$" <' <br> If ID == '>', ID :=' "Type III attr. $K$" >' <br> If ID == '=', ID := ' "Type III attr. $K$" =' |
| A Type III attr. value $N$ | If ID == '</>/=/≠/≥/≤', then <br> ID := ' "Type III attr." </>/=/≠/≥/≤ $N$' |
| Below, fewer, less, lower, max, most, smaller, < | If ID == 'Type III attr.', then <br>   ID := ' "Type III attr." <' <br> Else ID := '<' |
| Above, greater, higher, least, min, >, | If ID == ' "Type III attr." ', then <br>   ID := ' "Type III attr." >' <br> Else ID := '>' |
| Equal(s), = | If ID == 'Type III attr.', then <br>   ID := 'Type III attr. =' <br> Else ID := '=' |
| Newest, latest, | ID := 'group by year DESC' |
| Oldest, earliest, | ID := 'group by year' |
| Cheapest, inexpensive, | ID := 'group by price' |
| Lowest | ID := 'group by' |
| Between, range, within | If ID == 'Type III attr.', then <br>   ID := ' "Type III attr." between' <br> Else ID := 'between' |
| Other keyword | ID := '' |

*4.2.2 Processing incomplete questions*

To fully automate CQAds and avoid employing the user's feedback strategy to minimize the user's workload, CQAds analyzes *incomplete* questions and performs the "best guess" of the user's information need. If a numerical value $V$ (in an incomplete user's question) is not associated with a *keyword* that identifies the entity $V$ *quantifies*, CQAds considers $V$ as a potential value of each numerical attribute in the ads domain $D$ to which the user's question belonged. This is because among all the DB attributes in the schema defined for $D$, the numerical attributes are the only ones assigned numerical values and posted as a condition. CQAds excludes any record (answer) that does not include $V$ in the valid range of any of its Type III attributes (as defined in Section 4.3.2), which is determined by the smallest (largest, respectively) value under the pretended column of $V$ in the DB table of $D$. For example, in the car-ads domain, if $V$ does not fall into the range of 1985 and 2011, $V$ is treated as either $Price$ or $Mileage$, which are assumed to be the only other Type III attributes, and CQAds creates a SQL subquery that unions both possible selection conditions.

EXAMPLE 3. Consider the questions $Q$, "Honda accord 2000" and $Q'$, "Honda accord less than 4000", in which important information that determines the user's information need is missing. In the case of $Q$, CQAds interprets 2000 as Year, Price, or Mileage, since 2000 is in the range of year, price, and mileage of cars. In the case of $Q'$, however, 4000 is treated as Price or Mileage, but not Year, since 4000 is not in the range of valid years for cars □

*4.2.3 Shorthand notations*

Users' questions tend to vary in how they refer to data values. Regarding a car with four doors, any of the expressions '4dr', '4 dr', 'four door', '4 doors', '4-door', or '4doors' could be used. We developed a simple, yet effective, perl script that automatically detects the variance of "shorthand notation" of a data value. The perl script is based on the fact that any shorthand notation $N$ of a data value $V$ only includes *characters* from $V$, and the characters in $N$ should have the *same* order as characters in $V$. Hence, when a user specifies a data value $A$ in a question, a DB record $R$ is considered relevant with respect to $A$ if (i) an exact match for $A$ is found in $R$, (ii) $A$ is a detected shorthand notation of a data value in $R$, or (iii) there exist a shorthand notation of $A$ in $R$. Experiments on 1,000 ads in various domains show that our Perl script achieves a 98% accuracy in detecting shorthand notations.

### 4.3 Non-Boolean questions evaluation process

In evaluating the selection criteria specified in a non-Boolean question, *superlatives* are considered *after* all the other search criteria. This evaluation strategy yields the correct answers to the corresponding question, since the results retrieved by evaluating superlatives are DB records with the $max/min$ attribute values on which they are evaluated. Consider the question "(Find the) cheapest Honda". If the cheapest Hondas are more expensive than the cheapest Toyotas, evaluating 'cheapest' first retrieves only Toyotas, upon which searching for 'Honda' yields no results, which is incorrect. On the other hand, evaluating 'Honda' first, followed by 'cheapest,' yields the cheapest Hondas in the answer set.

Even though superlatives should be evaluated last, the order in which other conditions are evaluated does not change the end results, since they are evaluated in a commutative AND order. However, in order to speed up the evaluation of (Boolean) ads questions processed by CQAds, it is required that

1. Type I attribute values are evaluated first, since each of their corresponding attributes is defined as the primary-indexed field in its relation schema.

2. Type II attribute values, if there are any, are evaluated on the set of records extracted in Step 1, since each Type II attribute is defined as a secondary-indexed field.

3. Boundaries on Type III attributes and their values, if they exist, are evaluated on the records retrieved in Step 2.

4. Superlatives, if they exist, are evaluated on the records created in Step 3 which yield the answers to the given question.

*4.3.1 Partial matching of selection conditions*

Occasionally, the selection criteria specified in a question $Q$ only partially matches the corresponding values in a data record (i.e., an ad) or retrieves only a few matched records. To enhance the quality and quantity of the retrieved results under this scenario, CQAds relaxes the selection criteria specified in $Q$ so that other closely related ads can be extracted. Thus, for a question with $N$ ($\geq 2$) conditions, CQAds removes each condition in turn, using the remaining $N$-1 conditions in creating multiple queries, and retrieves the results of each modified query. For example, if a user is seeking a '2-door car for less than $6000', CQAds first retrieves cars that are 2-door under $6000, followed by any ads with '2-door' and price 'less than $6000' individually. For questions with *one* condition $C$, CQAds applies the similarity-matching strategy to retrieve records that partially satisfy $C$, if needed (see details in Section 4.3.2).

Practically, multiple conditions can be removed simultaneously, i.e., using up to $N$-2 or $N$-3 remaining conditions, etc. However, the more combinations of conditions to be considered, the *longer* the question processing time is required, and more importantly the *less* likely the results satisfy the users' requirements specified in their questions. Furthermore, according to the Search Engine User Behavior Study Report published by iProspect [9], 88% of Web search engine users only view the first three pages of search results (i.e., the first 30 results). Based on this statistical data, CQAds retrieves up to 30 (in)exact matched records for each question.

### 4.3.2 Similarity measures of attribute values

In designing the $N$-1 partial-matching strategy for enriching the quality and quantity of retrieved answers to an ads question $Q$, we have defined the *degree of similarity* of an attribute value $V$ in each record and the corresponding value $T$ in $Q$ that is excluded from the $N$-1 matching conditions for $ranking$ partially-matched results.

**Type I Values.** To measure the degree of similarity between a partially-matched Type I attribute value in an ads record and the corresponding one expressed in $Q$ belonged to an ads domain $D$, we consult the $TI$-$matrix$ for $D$. To construct the $TI$-$matrix$ for $D$, which includes the similarity values between any two distinct Type I attribute values in $D$, CQAds relies on query logs obtained from local ads search engines. Each query log includes a number of query sessions, each of which captures a period of sustained user activities on the corresponding ads search engine. Each log session differs in length and includes (i) a $user\ ID$, (ii) the $query\ text$, (iii) the $date$ and $time$ of each search, and (iv) optionally $clicked\ documents$. A $user\ ID$, which is an anonymous identifier of the user who performed the search, determines the boundary of each session (as each user ID is unique and associated with one session). The $query\ text$ are the keywords in a user query, and multiple queries can be created under the same session. The $date$ and $time$ of search can be used to determine the relative importance of the query results page based on the $time$ the user spent on that page, and $clicked\ documents$ are retrieved documents that the user has clicked on and are represented by their titles and ranked by the corresponding ads search engine.

The *similarity* value of any two distinct $Type$ I attribute values $A$ and $B$, denoted $TI\_Sim$, in the $TI$-$matrix$ is computed in Equation 3 based on the following features that can be determined using a query log:

(1) Number of times $A$ is modified to $B$ in the query log or vice versa, denoted $Mod(A, B)$.

(2) Average time between submissions of $A$ and $B$ in the same session, denoted $Time(A, B)$.

(3) Average time spent on an ad containing $B$ when $A$ is searched or vice versa, denoted $Ad\_Time(A, B)$.

(4) Ranking of an ad (as determined by the ads search engine that provides the query log) containing $B$ when $A$ is searched for, or vice versa, which is averaged over the entire query log, denoted $Rank(A, B)$. The *higher* $B$ is ranked, the *more likely* $B$ is similar to $A$ as determined by the corresponding search engine that ranks $B$.

(5) Number of times a document containing $B$ is clicked when $A$ is searched or vice versa, denoted $Click(A, B)$.

To normalize each of our feature values, we divide the resulting value of a feature $F$ by the maximum possible value of $F$ derived from our query log so that each factor value is in the range of [0..1].

$$TI\_Sim(A,B) = Mod(A,B) + Time(A,B) + Ad\_Time(A,B) \\ + Rank(A,B) + Click(A,B) \quad (3)$$

**Type II Values**. In establishing the *similarity* values among the *properties* of various products in different ads of the same domain, we use the $word$-$similarity$ matrix, denoted $WS$-$matrix$, introduced in [11]. $WS$-$matrix$ is a 54,625 × 54,625 symmetric matrix, which contains the similarity values of pairs of non-stop, stemmed words, i.e., words reduced to their grammatical root. The similarity value between any two non-stop, stemmed words $w_i$ and $w_j$ in $WS$-$matrix$ is computed by using the (i) *frequency* of *co-occurrence* and (ii) relative *distance* of $w_i$ and $w_j$ in a document. $WS$-$matrix$ was constructed using the documents in the Wikipedia document collection (en.wikipedia.org /wiki/Wikipedia:Databasedownload) that consists of 930,000 documents written by more than 89,000 authors on various topics, which are diverse in content and writing styles. The $WS\_matrix$ has been adopted successfully in solving various information retrieval problems [16]. The similarity between an attribute value $T$, such as 'white', in a question and an attribute value $V$, such as 'blue', in a data record, denoted $Feat\_Sim(T, V)$, is a similarity value in $WS$-$matrix$.

**Type III Values**. In establishing the *similarity* among *numerical attribute values* in different ads domains, we apply Equation 4.

$$Num\_Sim(T, V) = 1 - \frac{|T - V|}{Attribute\_Value\_Range} \quad (4)$$

where $Attribute\_Value\_Range$ is the normalization factor of $Num\_Sim$, which is different for each numerical-value attribute in the DB, and the complement is applied so that the *closer* $T$ and $V$ are, the *higher* their $Num\_Sim$ value is.

The $Attribute\_Value\_Range$ of each Type III attribute $A$ in an ads domain is determined by using the statistical data extracted from ebay.com. We obtain from ebay.com the 10 highest (lowest, respectively) values for $A$ and subtract the averaged *minimum* values of $A$ from the averaged *maximum* values of $A$ to establish the attribute value range for $A$,

EXAMPLE 4. Consider question $Q$ "Find all $10,000 cars". A car ad that includes a price of $11,000 is closer to the price specified in $Q$ than another car ad which includes a price tag of $7,500, since $Num\_Sim$($10,000, $7,500) = 1 - 2,500/10,000 = 0.75 $\leq$ $Num\_Sim$($10,000, $11,000) = 1 - 1,000/10,000 = 0.90, assuming that 10,000 is the price range for cars determined by the 10 maximum (minimum, respectively) values extracted from ebay.com. □

The *ranking similarity* value, $Rank\_Sim$, between a user's question $Q$ and a partially-matched record $r$ in which $T$ and $V$ occur, respectively is computed as follows:

$Rank\_Sim(r, Q) = (N - 1) +$

$$\begin{cases} TI\_Sim(T,V) & \text{if } T \text{ and } V \text{ are Type I attribute values} \\ Feat\_Sim(T,V) & \text{if } T \text{ and } V \text{ are Type II attribute values} \\ Num\_Sim(T,V) & \text{if } T \text{ and } V \text{ are Type III attribute values} \end{cases}$$
(5)

where $N$ is the number of attributes specified in $Q$. $Num\_Sim$ is in the range of [0..1], whereas $TI\_Sim$ and $Feat\_Sim$ are in different numerical scales. Hence, $TI\_Sim$ and $Feat\_Sim$ are normalized by the maximum possible value in each of their corresponding matrices. The value $N$-1 is added to account for the *exact* matches between the corresponding attribute values in $Q$ and $r$, each of which is assigned the value of 1.

EXAMPLE 5. Table 2 shows a portion of the partially-matched answers to the question $Q$ "Find Honda Accord blue less than 15,000 dollars" that CQAds retrieves and ranks using Equation 5. Table 2 also includes the partially-matched attribute value, which is highlighted, in each record, in addition to the similarity-measure strategy adopted for computing the ranking. □

## 4.4 Boolean questions processing

Occasionally, an ads question includes Boolean operators: AND, OR, NOT. We classify a user's question into one of the two categories that involve Boolean operators: implicit Boolean questions and explicit Boolean questions. *Explicit Boolean questions* are questions which contain at least one AND or one OR, whereas *implicit Boolean questions* contain neither, but include at least one negated attribute value[1] or two mutually-exclusive attribute values[2]. In our QA system, mutual exclusion applies only to Types I and II attribute values, since compatible Type III attribute values are combined (see details in Section 4.4.1). Analyzing implicit and explicit Boolean questions, which make up almost one-fifth of the ads questions created by Facebook users in various surveys conducted by us for verifying the effectiveness of CQAds, is not a trivial task, since these Boolean questions are not necessarily well-formed and may be ambiguous. As a result, their information needs may not be precisely stated. In this section, we define a set of rules for interpreting the (potential) information needs in Boolean questions.

### 4.4.1 Evaluating implicit Boolean questions

Users of QA systems may include in their questions (i) negated attribute values without explicitly specifying the Boolean operator NOT, e.g., "Any car except a blue one", or (ii) attribute values that cannot co-exist, e.g., "blue, red Toyota", which are mutually-exclusive values that describe the same property of a car without any OR operator[3] explicitly specified in between. To address these non-trivial problems, CQAds applies the following set of combination rules in processing implicit Boolean selection criteria, which are evaluated in a left-to-right manner, in an ad question:

(1) For each group of values that are valid domain values of a Type III attribute, do

  (a) If any value is negated, the negated quantifier is replaced by its complement.

  (b) If two or more values are specified with the keywords or their synonyms (as shown in Table 1) "less than" ("more than", respectively), "equal", or any combination of the two, then the lower (higher, respectively) of the two quantified values is retained and the other is discarded.

  (c) If one of the values is specified with a "less than", "equal", "less than or equal to", or their synonyms, and another with a "greater than", "equal", "greater than or equal to", or their synonyms, then the two quantified values are combined with the keyword "between", unless the two quantified values do not overlap. In the latter case, CQAds displays the message "search retrieved no results" and terminates the question-evaluation process. This process is repeated by combining any intermediate results with a remaining value.

(2) Evaluate each sequence of consecutive, (non-)negated Type II, denoted $T_2$, such that

  (a) Negated attribute values in $T_2$ are ANDed together, whereas non-negated ones are ORed if they are mutually exclusive; otherwise, they are ANDed together.

  (b) Each subexpression created in Rule 2a is ANDed with the closest (negated) Type I attribute value $A$, if $A$ exists. $A$ is associated with only one subexpression.

(3) Repeat Step 2 for Type III attribute values or combined Type III attribute values generated in Step 1.

(4) If there are more than one subexpression created in Step 2 or 3 that includes a (negated) Type I attribute value, these subexpressions are ORed together.

Rule 2a (3a, respectively) identifies *which* and *how* Type II (Type III, respectively) attribute values in a question should be combined, whereas Rule 2b (3b, respectively) dictates *how* the newly created subexpressions can be combined with a Type I attribute value, if it exists. We treat a Type I attribute value as a main searching criteria of a question and each sequence of Type II/Type III attribute values are "right-associated" with a Type I attribute value $A$, since they are treated as "descriptive properties" of $A$, if $A$ exists.

EXAMPLE 6. Given the implicit Boolean question $Q_1$, "Any car priced below $7000 and not less than $2000", CQAds applies Rule 1a to transform "not less than $2000" to "more than or equal to $2000". Thereafter, "below $7000" and "more than or equal to $2000" are combined as "between $2000 AND less than $7000" using Rule 1c.

Consider another question $Q_2$, "I want a Toyota Corolla or a silver not manual not 2-dr Honda Accord". Applying Rule 2a on "not manual" and "not 2-dr", which are consecutive, negated Type II attribute values, yields the subexpression "NOT manual AND NOT 2-dr". The subexpression is ANDed with "silver", another consecutive Type II attribute value. Using Rule 2b, "silver AND NOT manual AND NOT 2-dr" is "right-associated", i.e., ANDed, with "Honda AND Accord". Using Rule 4, the subexpressions "silver AND NOT manual AND NOT 2-dr AND Honda AND Accord" and "Toyota AND Corolla" are combined with an OR, since each subexpression contains a (mutually exclusive) value of the same Type I attribute, i.e., Make (Model, respectively). Both $Q_1$ and $Q_2$ are questions created by Facebook users (see Section 5.4). □

### 4.4.2 Evaluating explicit Boolean questions

In the car-ads and domain-specific question Facebook surveys (see Section 5.1) in which the users were asked to submit sample questions, we found that out of 650 submitted questions, only 34 (5.2%) were explicit Boolean questions. Moreover, as claimed by Ross and Wolfram [19], only between 3 to 5% of web searches include Boolean operators, which matches the percentage of explicit Boolean questions created by Facebook users who participated in our surveys. Furthermore, interpreting the exact information need expressed in an arbitrary Boolean question/query without an explicit order specified for evaluating its subexpressions is a very difficult, if not impossible, task (see discussions in [14]). For these reasons, we did not develop a new set of evaluation rules for processing any explicit Boolean ads question $Q$. Instead, CQAds excludes all the Boolean ANDs and ORs from $Q$ and evaluates $Q$ as an *implicit* Boolean or non-Boolean question, depending on the question. In special cases, when $Q$ consists of a sequence of attribute values of any types separated by only ORs (ANDs, respectively), $Q$ is evaluated as is (without ANDs, respectively).

---

[1] A negation is specified in a question $Q$ if one of the following keywords (or their stemmed versions) is detected in $Q$: not, no, without, except, excluding, remove, nothing, and leave out.

[2] Mutually-exclusive attribute values are domain values of the same attribute that cannot co-exist in an ads question.

[3] CQAds combines consecutive non-mutually-exclusive attribute values in an implicit question using logical ANDs by default.

**Table 2: Top-5 ranked, partial-matched answers to question $Q$, "Find Honda Accord blue less than 15,000 dollars", in Example 5**

| Ranking | Make | Model | Price | Features | $Rank\_Sim$ | Similarity Measure Used |
|---|---|---|---|---|---|---|
| 1 | **Chevy** | Malibu | 5899 | blue, anti-lock brake, power steering | 3.73 | $TI\_Sim$ on Make and Model |
| 2 | Toyota | **Camry** | 8561 | 4 cylinder, automatic, blue, . . . | 3.43 | $TI\_Sim$ on Make and Model |
| 3 | **Ford** | Focus | 6795 | blue, cd player, radio, power door locks, . . . | 2.91 | $TI\_Sim$ on Make and Model |
| 4 | Honda | Accord | **16536** | 4 cylinder, blue, 2 wheel drive, cassette player, . . . | 2.63 | $Num\_Sim$ on Price |
| 5 | Honda | Accord | 6600 | GPS system, **gold**, auto-off headlights, . . . | 2.31 | $Feat\_Sim$ on Color |

Even though CQAds treats explicit Boolean questions as either implicit or non-Boolean questions, we have observed that the performance of CQAds in not significantly affected based on the empirical study presented in Section 5.4. CQAds achieves 90% accuracy in correctly interpreting the information needs expressed in explicit Boolean questions by applying the evaluation rules of implicit Boolean questions on them.

### 4.5 Question execution

In this phase of the question-answering process, CQAds concatenates the search criterion (i.e., generated sub-queries) to create an SQL query statement. The SQL statement may have one or more *select criterion* (*sub-queries*) created from the corresponding user's question $Q$. The created statement logically AND all the sub-queries, if the keywords 'not' and 'OR' (or their synonyms) are not included in $Q$; otherwise, the sub-queries are concatenated using the previously defined Boolean rules (in Section 4.4.1).

EXAMPLE 7. Given the question $Q$, "Do you have automatic blue cars?", CQAds creates the following SQL query statement[4]:

SELECT * from $Car\_Ads$ WHERE Car_ID IN
 (SELECT Car_ID from $Car\_Ads$ $C$
  WHERE $C.Transmission$ = 'Automatic') **AND** Car_ID IN
 (SELECT Car_ID from $Car\_Ads$ $C$
  WHERE $C.Color$ = 'blue') □

The SQL query statement is then exported to the MySQL database engine, which is chosen for CQAds, to be processed and the results are presented to the user as answers to $Q$. Each answer retrieved is based on either exact or ranked, partial match. The answers are displayed on an HTML interface in a tabular manner.

Besides retrieving exact and/or partially-matched answers to a user's question $Q$, a major design issue of CQAds is the *efficiency* of processing $Q$. We have implemented a primary MySQL *substring index* [7] of length 3 on all the attributes of different ads domains in a MySQL DBS. Substring indexes are *shorter* than their corresponding entire column values, require *less* disk storage, and hold *more* keys in the cache memory for searching.

### 4.6 Adding a new ads domain

As the majority of the processing steps in creating a new ads domain $A$ are fully-automated it requires approximately $2\frac{1}{2}$ hours of manual labor to add $A$ to CQAds, for the remaining semi-automated steps detailed in Sections 4.1 and 4.2. Approximately 2 hours of manual labor are spent on verifying the correctness in which attribute values in ads (employed for constructing domain-specific tables) are annotated.

## 5. EXPERIMENTAL RESULTS

In this section, we describe the process through which we obtained the test data and discuss the evaluation metrics used for assessing the performance of CQAds. Hereafter, we report the experiments conducted on CQAds and analyze their results. To the best of our knowledge there is *no* publicly available QA system on ads that evaluates (Boolean) questions and considers partial answer matching, which can be compared against CQAds.

### 5.1 Test data

Since there is no benchmark dataset available for evaluating the performance of an ads QA system, to obtain a representative test dataset for verifying the effectiveness of CQAds in retrieving exact- and/or partially-matched answers to users' questions, we have solicited ads questions on eight ads domains through the Internet using Facebook. The eight ads domains we consider are Cars, Motorcycles, Clothing, Computer Science Jobs, Furniture, Food Coupons, Musical Instruments, and Jewellery. These domains are diverse and representative of everyday living essentials, i.e., transportation, clothing, jobs, housing, food, and entertainment.

Facebook, which is a free-access social network, was chosen for conducting the surveys, since it reaches people from various age groups and backgrounds who can objectively perform the required evaluations. We prepared two surveys, the car-ads and domain-specific question surveys. The *car-ads* survey includes 4 questions, three requiring short answers and one in multiple choice format (Question 4). The domain-specific question survey, which includes only the first question in the car-ads survey, is repeated for each of the remaining seven ads domains. The surveys were sent out on June 1, 2009 to different Facebook users who were asked to forward the surveys to others. By March 25, 2010, we received 650 responses (80 for the car-ads survey and 570 for the domain-specific question survey). Figure 1 shows a response to the Facebook application that includes the car-ads survey questions and their answers.

The answers to the first question from both surveys yield the set of *test questions* used for our empirical study, whereas the answers to the second question from the car-ads survey show that 91% of surveyed users who do not find exact matches for their searches would likely remove (or modify) a feature from their original search, which is our $N$-1 *(partial) matching* approach. Among the answers to the fourth question, 93% percent of the users indicate that they would rather search cars with $similar$ features. Last, but not least, the third question is included for verifying the appropriateness of our choice of 30 as the ideal number of retrieved answers to a user's ad question, which is close to (the average) 26. The last three questions on the car-ad survey were not duplicated in the domain-ad survey, since the answers to these questions are expected to remain the same for the other ads domains.

---
[4]If no exact matches are found for $Q$, then "AND" (in bold) in the SQL query is replaced by "OR" to retrieve partial answers based on the "$N$-1" strategy.

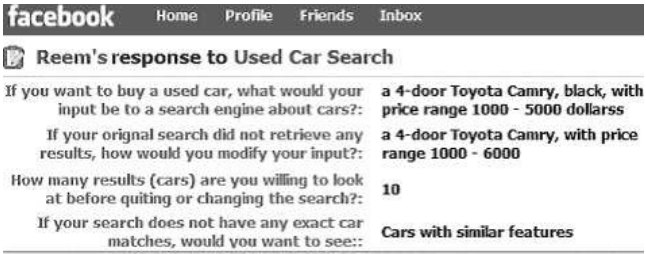

**Figure 1: A response to the car-ads Facebook survey.**

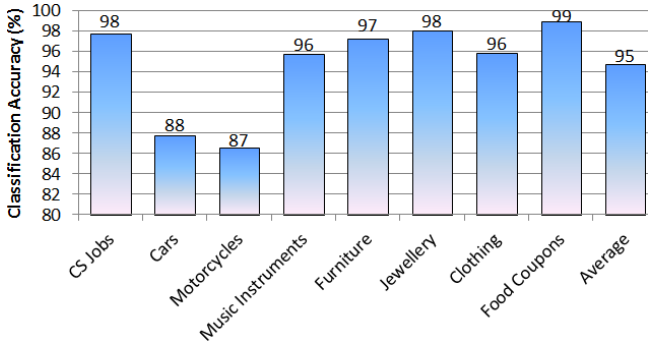

**Figure 2: Classification accuracy of the 650 ads questions.**

### 5.2 Performance analysis on question classification

To evaluate the effectiveness of CQAds in $classifying$ ads questions, we rely on the accuracy ratio as defined below.

$$Accuracy = \frac{Correctly\_classified\_instances}{Total\_number\_of\_instances} \quad (6)$$

where $Total\_number\_of\_instances$ is the total number of questions to classify, which is 650 in our case, and $Correctly\_classified\_instances$ is the number of questions correctly assigned to their corresponding domains by the Naive Bayes classifier.

Figure 2 shows the accuracy ratios of classifying the 650 ads questions into each of the eight ads domains previously introduced, as well as the average classification accuracy on the ads domains achieved by CQAds, which are in the (upper) ninety percentile. Ads in Cars-for-Sale and Motorcycles-for-Sale domains achieve the lowest accuracy ratios (in the upper eighty percentile) among the eight domains due to the existence of common keywords between the two domains.

### 5.3 Evaluation measures of retrieving exact-matched ads (answers)

To measure the performance of CQAds in retrieving answers that exactly match the constraints specified in users' questions, we considered the 650 responses to the first question in our two Facebook surveys and evaluated them using CQAds. The evaluation metrics for measuring the correctness of retrieving exactly-matched answers to an ads question are (i) $precision$ ($P$), which is the ratio of the number of $correct\ matches$ retrieved by CQAds over the total number of $records\ retrieved$ by SQL, (ii) $recall$ ($R$), which is the ratio of the number of $correct\ matches$ retrieved by CQAds over the number of $correct\ answers$ in the DB, and (iii) $F$-$measure$ = $\frac{2}{\frac{1}{P}+\frac{1}{R}}$, where a $correct\ match$ is a retrieved answer (up till the $30^{th}$), i.e., DB record, that satisfies all the search criteria specified in a question.

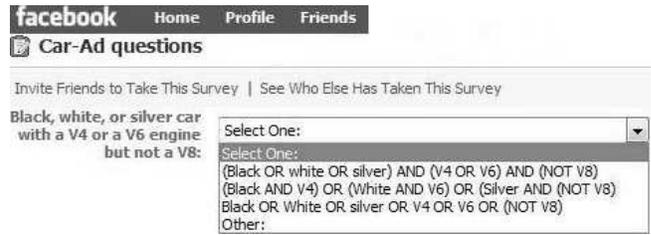

**Figure 3: A sampled question from the second Boolean Facebook survey with its interpretations.**

The averaged $precision$, $recall$, and $F$-$measure$ for the 650 test questions, yield 93.8% precision, 92.7% recall, and an F-measure of 93.2%. We have observed that most of the test questions yield 100% for $precision$ and $recall$, whereas a few yield 0%, i.e., answers are either correct or incorrect.

### 5.4 Evaluating the interpretations of implicit and explicit Boolean questions

As stated in Section 5.1, since there is no baseline measure or established dataset for evaluating the performance of a QA system on ads (needless to say on Boolean questions), we conducted another two Facebook surveys, a Boolean-question survey and a Boolean survey, to analyze the accuracy of the interpretations generated by CQAds on the intended information needs specified in Boolean ads questions. The *Boolean-question* Facebook survey requests the users to submit questions containing (i) at least one Boolean AND or OR, (ii) two mutually-exclusive elements, (iii) a negation, or (iv) any combination of (i), (ii), and (iii). The survey was posted between September 1, 2009 and December 2009, and we received 182 responses, i.e., Facebook-user-created Boolean questions. (All the Boolean questions created by Facebook users can be found under students.cs.byu.edu/˜rmq3/QAresults.) Out of these responses, 89 solicited questions are implicit and 93 explicit. We used the submitted questions from the Boolean-question survey for conducting the Boolean survey which verifies the accuracy of CQAds in interpreting the questions using the rules presented in Section 4.4.1.

To construct the second survey, the Boolean survey, we chose 10 questions out of the 182 responses to the Boolean-question survey, which are unique in terms of their information needs, the choices of Boolean/Negation operators, and the individual function of each operator in the question. Out of the 10 sampled questions, seven are *explicit* questions and three are *implicit*. For each of these questions, we included in the Boolean survey (i) the interpretation of the question generated by CQAds, (ii) two other manually-created interpretations, and (iii) an option that allows the user to enter his/her own interpretation, assuming that the user disagrees with any given interpretations. Figure 3 shows one of the 10 sampled questions and its interpretations, i.e., the corresponding Boolean expressions.

The Boolean survey was posted on December 2009, and up till March 25, 2010, we received 90 responses. (All the responses can be retrieved under students.cs.byu.edu/˜rmq3/QAresults.) To assess the $accuracy$ of the interpretations generated by CQAds, we calculated the $fraction$ of the number of responses on each sampled question that choose the CQAds' interpretation divided by the total number of responses for the sampled question. Figure 4 shows the accuracy obtained for each question, in which $Q_2$, $Q_3$, and $Q_4$ are the three implicit questions. CQAds achieves an average of 90.2% accuracy rate, which demonstrates that CQAds, although not based on user's feedback, is reliable in transforming the information need specified in a(n) (ambiguous) Boolean question. Furthermore, CQAds achieves an average of 90.3% accuracy on interpreting *implicit* questions and 90.1% on *explicit* questions, which shows that in the absence of a set of evaluation rules for handling *explicit* Boolean questions, CQAds achieves a high accuracy rate by evaluating them as *implicit* Boolean questions.

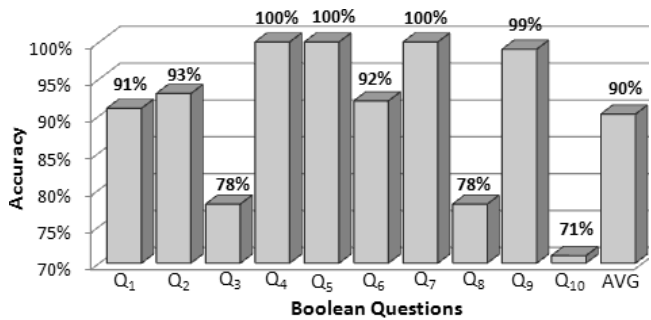

**Figure 4: Boolean question interpretation accuracy achieved by CQAds obtained using the second Boolean Facebook survey.**

As shown in Figure 4, the interpretations of $Q_3$, $Q_8$, and $Q_{10}$ performed by CQAds are not well-received by some users. Both $Q_3$, "Show me Black Silver cars", and $Q_8$, "Focus, Corolla, or Civic. Show only black and grey cars", include two consecutive mutually-exclusive attribute values that are not *OR*ed. CQAds changes the AND to OR, since logically ANDing two mutually-exclusive attribute values would yield no results. However, 22% of the Facebook users who participated in the Boolean survey believe that $Q_3$ (and $Q_8$, respectively) asks for car ads that include both attribute values. CQAds interprets $Q_8$ as "(Focus OR Corolla OR Civic) AND (Black OR grey)", whereas, 22% of the users interpret $Q_8$ as "(Focus OR Corolla OR Civic) AND (Black with grey)". A similar interpretation applies to $Q_3$ in which "Black Silver cars" is specified. As for $Q_{10}$, "Black Mustang with gps, exclude 2 wheel drive, or a yellow corvette without a gps", CQAds interprets it as "(not 2-wheel drive Black Mustang with gps) OR (yellow corvette without gps)", whereas 29% of the users believe "exclude" should be applied to "yellow corvette" as well.

Since Boolean operators are the same from one ads domain to the other, evaluating the interpretations of Boolean questions created by CQAds on the car-ads domain are the same for CQAds on other ads domains as well.

## 5.5 Evaluating the partially-matched answers ranking strategy

In this section, we present the evaluation on our ranking strategy which orders partially-matched answers to users' questions and compare it with other representative ranking approaches. Again, since there is no benchmark data available for evaluating the partially matched answers to users' questions on ads, we have used individual appraisers to assess the ranking strategy of CQAds and compare it with other state-of-the-art ranking approaches.

To gather the individual appraiser's ranking, we conducted another survey on Facebook beginning in June 2009, called *ranking survey*, which lasted till March 2010. In the survey, we provided 40 different randomly-selected questions from the first two Facebook surveys, denoted $Test\_Questions$, on the eight different ads do-mains (5 on each domain). For each question, we arranged in random order the top-5 partially-matched answers retrieved by each of the four ranking approaches[5] (presented in Section 5.5.2) that we compare CQAds against which yield 25 answers (5 answers x 5 ranking approaches) to each question. The appraisers were required to read a question $Q$ and determine which given answers are (un)related to $Q$. We requested Facebook users to evaluate only five answers to each question from each ranking approach, since examining the relevance of answers to questions is a time-consuming task. Overall, 886 responses to the survey were collected.

### 5.5.1 Evaluation metrics

To measure the effectiveness of the ranking strategy of CQAds on partially-matched answers to questions (introduced in Section 4.3.2), we apply two well-known information retrieval metrics, the (overall) *Precision at K* and *Mean Reciprocal Rank* [5].

The $P@K$ measures the overall user's satisfaction with the top-$K$ ranked answers (generated by CQAds) to a particular question in $Test\_Questions$.

$$P@K = \frac{\sum_{i=1}^{N} \frac{Number\_of\_Related\_Answers_i}{K}}{N} \quad (7)$$

where $K$ is the (pre-defined) number of answers to be considered, $N$ is the total number of questions in $Test\_Questions$, which is 40, $i$ is the $i^{th}$ question in $Test\_Questions$, and $Number\_of\_Related\_Answers_i$ is the average number of answers (out of $K$) that are treated as *related* to the $i^{th}$ question by the appraisers who evaluated question $i$. Note that in our study, we set $K$ to be 1 and 5 to evaluate the relatedness of the answers positioned at the *top* and *overall* in the ranking, respectively.

Besides $P@K$, we further evaluate the ranking strategy of CQAds using the *Mean Reciprocal Rank* ($MRR$) metric. $MRR$ is the averaged sum of the reciprocal of the ranking position of the *first* related answer among the top-5 answers, if there is any, or 0, otherwise for each question in $Test\_Questions$.

$$MRR = \frac{1}{N} \sum_{i=1}^{N} \frac{1}{r_i} \quad (8)$$

where $r_i$ is the average (position in the) rank of the *first related* answer to question $i$ in $Test\_Questions$, if it exists; otherwise, $r_i = \infty$, and $N$ and $i$ are as defined in Equation 7.

$P@K$ and $MRR$ evaluate the ranking strategy of CQAds, such that the *higher* relevant answers are positioned in the ranking list, the *higher* their corresponding $P@K$ and $MRR$ scores are.

### 5.5.2 Ranking methods used for comparison purpose

To assess the effectiveness of CQAds in ranking partially-matched answers to a user's question, we compare its performance, in terms of $P@1$, $P@5$, and $MRR$, with four other ranking approaches: (i) Random [13], (ii) cosine similarity [12], (iii) AIMQ [15], and (iv) FAQFinder [3].

*Random* ranking, which presents the partially-matched answers to a question in a random order, provides a *baseline* to determine how well a ranking approach can meet the user's expectations in terms of ordering the answers with various degrees of relevance to a question. The *cosine similarity* approach, which is used in the Vector Space Model, computes a similarity value between two vectors (in our case, one vector representing a user's question $Q$ and the other a partially-matched answer $A$ to $Q$) by measuring the

---
[5]The random-ordered answers avoids imposing bias on the appraisers regarding the positions in the ranking of the retrieved answers.

226

angular distance between them. In our implementation, the cosine similarity between $Q$ and $A$ is computed using binary weights such that for each selection constraint $C$ specified in $Q$, '1' represents the satisfaction of $C$ by $A$, and '0' otherwise. The *higher* the computed cosine-similarity value of $A$ is, the *higher* the ranking of $A$ with respect to $Q$ is.

$AIMQ$ relies on attribute-value pairs (denoted AV-pairs) to generate the associated supertuple of each attribute. A supertuple is an inferred DB tuple that contains a set of attribute values, each of which includes a summary of values in the corresponding table column, and is used for calculating the similarity of categorical attributes. AIMQ determines the similarity between $Q$ and $A$ using Equation 9.

$$Sim(Q, A) = \sum_{i=1}^{n} W_{imp}(A_i) \times$$

$$\begin{cases} VSim(Q.A_i, A.A_i), & \text{if } Domain(A_i) = Categorical \\ 1 - \frac{|Q.A_i - A.A_i|}{Q.A_i}, & \text{if } Domain(A_i) = \text{Numerical} \end{cases}$$
(9)

where $n$ is the number of attributes in $Q$, $A_i$ in an attribute (in either $Q$ or $A$), $Wimp(A_i)$ is the importance weight of $A_i$, which in our implementation of AIMQ is set to be $\frac{1}{n}$ for each attribute, $Q.A_i$ ($A.A_i$, respectively) is the value of attribute $A_i$ in $Q$ ($A$, respectively), $1 - \frac{|Q.A_i - A.A_i|}{Q.A_i}$ generates the *similarity* between any two numerical attributes $Q.A_i$ and $A.A_i$ in which $|Q.A_i - A.A_i|$ is the absolute difference between the two numerical attribute values $Q.A_i$ and $A.A_i$, and $VSim(Q.A_i, A.A_i)$, which is the similarity between two categorical attributes $Q.A_i$ and $A.A_i$, is computed using Equation 10.

$$VSim(Q.A_i, A.A_i) = \sum_{i=1}^{n} J(C_1.A_i, C_2.A_i) \qquad (10)$$

where $A_i$ and $n$ are as defined in Equation 9, $C_1.A_i$ ($C_2.A_i$, respectively) is the supertuple of $Q.A_i$ ($A.A_i$, respectively), and $J(C_1.A_i, C_2.A_i)$ is the *Jaccard Coefficient* computed as $|C_1.A_i \cap C_2.A_i|/|C_1.A_i \cup C_2.A_i|$, which is the proportion of attribute values in common between $C_1$ and $C_2$ for $A_i$ and the distinct number of attribute values in $C_1$ and $C_2$.

In implementing $FAQFinder$, we (i) compute the weights for the TF-IDF similarity measure based on all the ads records in our DB, (ii) treat each ads data record in the DB as a document, and (iii) treat each question submitted by the user as a FAQ and each DB record to be ranked as in a FAQ file. (A detailed discussion on FAQFinder can be found in Section 2.2.)

### 5.5.3 Comparison results of our ranking approach

We computed the $P@1$, $P@5$, and $MRR$ values for each of the ranking approaches $R$ discussed in Section 5.5.2 using the total 886 responses provided by the Facebook users on ranked answers to each one of the 40 sampled questions in $Test\_Questions$. The $P@1$, $P@5$, and $MRR$ results are shown in Figure 5.

As illustrated in Figure 5, CQAds outperforms the other ranking approaches, which further verifies the *effectiveness* of the strategy adopted by CQAds for ranking partially-matched answers. A high $P@1$ score implies that the ranking strategy of CQAds is highly effective in presenting first answers that users are interested in. The *higher* $P@5$ score achieved by CQAds than other ranking approaches demonstrates that, in general, answers relevant to a question $Q$ are positioned *higher* by CQAds in the list of partially-matched answers retrieved in response to $Q$. Finally, the *higher* $MRR$ score achieved by CQAds than the other ranking approaches indicates that CQAds users browse through *less* partially-matched answers before locating the ones relevant to his/her question.

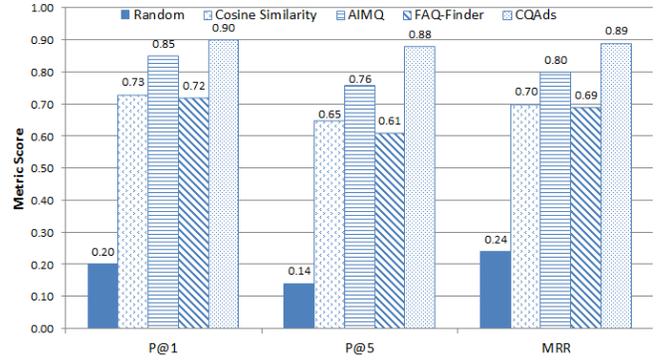

**Figure 5:** $Precision@K$ ($K$ = 1, 5) and $MRR$ scores on the (top-5) answers achieved by CQAds and other ranking approaches for the 40 questions in $Test\_Questions$.

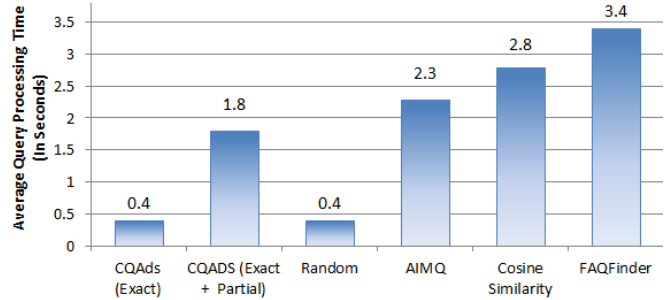

**Figure 6:** Average query processing time of CQAds and others.

We observed that the corresponding $P@1$, $P@5$, and $MRR$ values for FAQFinder are the *lowest* except the Random approach, since FAQFinder uses a simple method that does not compare numerical attributes. On individual domain level, we observed that the lowest scores on $P@1$, $P@5$, and $MRR$ for CQAds occur in the CS jobs ads domain. For this domain, appraisers did not rank the answers based on their similarity to the original question. For example, a C++ software programmer job is closely related to a C programmer job, but the appraisers ranked the answers based on which result is more relevant to their own expertise and experience, which is different from one user to another.

### 5.6 The efficiency of CQAds and other ranking approaches

The *efficiency* of CQAds and four other ranking approaches are presented in Figure 6, which shows the average query processing time for the 650 questions obtained from the car-ads and domain-question Facebook surveys. CQAds outperforms all ranking approaches except the Random ranking strategy. This is due to the fact that the Random approach does not perform any processing or employ any similarity measures, but rather selects records randomly, which is done fast. Unlike the compared ranking methodologies, CQAds retrieves exact matches first then partially-matched answers, if needed. Hence, there is an additional time imposed on CQAds for retrieving partially-matched answers. However, as illustrated in Figure 6, the query processing time for CQAds is the fastest compared to AIMQ, cosine similarity, and FAQFinder when partially matched and exact answers are retrieved.

## 6. CONCLUSIONS

We have introduced CQAds, a closed-domain question-answering (QA) system for advertisements (ads for short), which retrieves exact, as well as partially matched, answers to natural language questions on ads. CQAds uses an elegant approach to determine the *evaluation order* of selection criteria specified in a(n) (Boolean) ads question $Q$. Whenever CQAds retrieves a few or no *exact-matched* answers to $Q$, it extracts answers that partially match the selection criteria specified in $Q$ and *ranks* the answers based on a similarity-matching approach which relies on word-correlation factors as well as other domain-specific correlation matrices. We have also introduced a simple, yet effective set of rules for handling implicit/explicit *Boolean questions*, which enhance the functionality of our QA system. Furthermore, we have proposed a method for detecting *shorthand notations* and matching them to their corresponding original attribute values.

CQAds is a contribution to the QA community, since it introduces (i) the use of the *trie* structure to simply, yet effectively, determine missing information in a user's questions and correct any spelling mistakes, (ii) a novel similarity formula for determining the relevance of *partially-matched records* to a user's question, (iii) a fully automated approach to construct *similarity matrices* of categorical and numerical attributes independent of the domain of an ads question, (iv) a novel set of rules for interpreting the users' *intention* in (implicit) Boolean questions, and (v) a new approach for *transforming* users' questions into SQL queries. More importantly, CQAds provides a *framework* for answering questions on ads that does not require any specific domain algorithm and thus can easily be extended to answer questions on any ads domains, other than the eight domains evaluated in this paper.

Based on the results of the conducted empirical studies, CQAds achieves a 93.2% F-measure in retrieving answers that satisfy the users' information needs specified in ads questions. Moreover, CQAds obtains a 90% accuracy rate in interpreting the evaluation order of subexpressions in an ads question with implicit or explicit Boolean operators. CQAds outperforms well-known ranking approaches, in terms of the $P@K$ and $MRR$ metrics, which verifies that CQAds is reliable in ranking closely related answers that partially-match the search criteria specified in an ads question.

Although the percentage of Boolean questions is significantly lower than its counterpart, we plan to develop a set of well-defined evaluation rules to properly handle explicit Boolean ads questions as part of our future work. We also plan to study (i) the adaptation of CQAds to other domains, besides ads, (ii) automated database schema generation, (iii) the use of transformation rules to enhance the accuracy of matching records to questions, and (iv) de-duplication of data to remove similar data records from a DB.

## 7. REFERENCES

[1] B. Allison. An improved hierarchical bayesian model of language for document classification. In *COLING*, pages 25–32, 2008.

[2] M. Bilotti, P. Oglivie, J. Callan, and E. Nyberg. Structured retrieval for question answering. In *ACM SIGIR*, pages 351–358, 2007.

[3] R. Burke, K. Hammnod, V. Kulyykin, S. Lytinen, N. Tomuro, and S. Schoenberg. Question answering from frequently-asked question files: Experiences with the faq finder system. *AI Magazine*, 18(2), 1997.

[4] H. Chung, Y. Song, K. Han, D. Yoon, J. Lee, and H. Rim. A practical qa system in restricted domains. In *ACL Workshop on QA in Restricted Domains*, pages 39–45, 2004.

[5] W. Croft, D. Metzler, and T. Strohman. *Search Engines: Information Retrieval in Practice*. Addison Wesley, 2010.

[6] G. Das, V. Hristidis, N. Kapoor, and S. Sudarshan. Ordering the attributes of query results. In *ACM SIGMOD*, pages 395–406, 2006.

[7] P. Dubois. *MySQL*, $3^{rd}$ *Ed.* Pearson Education, 2005.

[8] O. Ferrandez, R. Izquierdo, S. Ferrandez, and J. Vicedo. Addressing ontology-based question answering with collections of user queries. *IPM*, 45(2):175–188, 2009.

[9] iProspect. iprospect.com/premiumPDFs/WhitePaper_2006_SearchEngineUserBehavior.pdf, 2006.

[10] W. Kießling. Foundations of preferences in database systems. In *VLDB*, pages 311–322, 2002.

[11] J. Koberstein and Y.-K. Ng. Using word clusters to detect similar web documents. In *KSEM*, pages 215–228, 2006.

[12] C. Manning, P. Raghavan, and H. Schutze. *Introduction to Information Retrieval*. Cambridge University Press, 2008.

[13] X. Meng, Z. Ma, and L. Yan. Answering approximate queries over autonomous web databases. In *WWW*, pages 1021–1030, 2009.

[14] D. Moldovan, M. Paşca, S. Harabagiu, and M. Surdeanu. Performance issues and error analysis in an open-domain question answering system. *TOIS*, 21(2):133–154, 2003.

[15] U. Nambiar and S. Kambhampati. Answering imprecise queries over autonomous web databases. In *ICDE*, pages 45–54, 2006.

[16] M. Pera, W. Lund, and Y.-K. Ng. A sophisticated library search strategy using folksonomies and similarity matches. *JASIST*, 60(7):1392–1406, 2009.

[17] M. Pera, R. Qumsiyeh, and Y.-K. Ng. Web-based closed-domain data extraction on online advertisements. Submitted to *Journal of Information Systems*, July 2011.

[18] L. Pizzato and D. Molla. Indexing on semantic roles for question answering. In *COLING*, pages 74–81, 2008.

[19] N. Ross and D. Wolfram. End-user searching on the internet: An analysis of term pair topics submitted to the excite search engine. *JASIS*, 51(10):949–958, 2002.

[20] R. Terol, P. Martinez-Barco, and M. Palomar. A knowledge based method for the medical question answering problem. *Computers in Biology and Medicine*, 10:1511–1521, 2007.

[21] M. Vargas-Vera and M. Lytras. Aqua: A closed-domain question answering system. *ISM*, 27(3):217–225, 2010.

[22] D. Wang. A domain-specific question answering system based on ontology and question templates. In *ACIS SNPD*, pages 151–156, 2010.

[23] Z. Wang and X. Luo. A semantic pattern for restricted domain chinese question answering. In *IMCLC*, pages 1333–1338, 2009.